\documentclass[12pt]{article}

\usepackage{epsf}
\usepackage[dvips]{graphicx}

\begin{document}

\begin{center}
{\bfseries SCALING BEHAVIOUR  OF EXCLUSIVE  REACTIONS 
   WITH THE DEUTERON AND  $^3He$ AT HIGH $p_T$ IN THE GEV REGION}


\vskip 5mm

Yu.N. Uzikov$^{1 \dag}$

\vskip 5mm

{\small
(1) {\it
Joint Institue for Nuclear Researches,LNP, Dubna, Moscow reg. Russia 
}

$\dag$ {\it
  E-mail: uzikov@nusun.jinr.ru
}}
\end{center}

\vskip 5mm

\begin{center}
\begin{minipage}{150mm}
\centerline{\bf Abstract}

 The  scaling  behaviour $s^{-11}$ of the cross section of the reaction 
 $\gamma d\to pn$  observed at SLAC and Jlab 
 at energies $E_{\gamma}=1-4$ GeV  and large $p_T$ most likely
 displays quark degrees of freedom in the deuteron. We show
 that the cross sections of  the  $dd\to~^3Hp$ and $pd\to ^3Hen$
  reactions measured at
 SATURNE  follow  the scaling  regime $s^{-22}$  
 at $T_d= 0.5 -1.2$ GeV and  $\theta_{cm}=50^\circ-60^\circ$.
 A necessity to get new data on this  
and other exclusive reactions is argued.
\end{minipage}
\end{center}

\vskip 10mm

\section{Introduction}
 Search for transition region from hadron
 to quark-gluon degrees of freedom
 in nuclear structure at short distances between nucleons ($r_{NN} < 0.5$~fm) 
 is an important   problem of particle and nuclear physics.
 This transition is expected to occur in processes at high 
 transferred momenta allowing to probe  dense fluctuations of
 nuclear matter (fluctons)
 in nuclear structure \cite{blokhintsev,baldin71}.
 So, very interesting scaling features  were found  in inclusive spectra of
 deep inelastic nuclear reactions in  {\it cumulative} region.
 In review \cite{Leksin}  these features are interpreted as a
 manifestation of ``drops'' of the quark phase
 in nuclei.

 Another possible signature for this transition is related with
 the  constituent counting rules (CCR) \cite{matveevmt}. According to
 dimensional  scaling \cite{matveevmt}
 the differential cross
 section  of a binary reaction $AB\to CD$ at  high enough
 incident energy  can be parameterized for a given c.m.s. scattering angle
 $\theta_{cm}$  as
\begin{equation}
\label{general}
\frac{d\sigma}{d\,t}(AB\to CD)= \frac{f(t/s)}{s^{n-2}},
\end{equation}
where 
 $n=N_A+N_B+N_C+N_D$ and $N_i$ is the minimum number of 
 point-like constituents in the {\it i-th} hadron (for a lepton and photon 
 one has $N_l=1$),
 $f(s/t)$ is a function of  $\theta_{cm}$.
 The CCR follows from perturbative QCD (pQCD).
 Recently  the similar
 scaling behaviour was found within  nonperturbative theories which are
 dual to QCD \cite{polchinski}. Existing high energy  data  for 
 many  measured  hard  processes
 with free hadrons appear to be consistent with the CCR
\cite{white}.

\begin{figure}[h]
 \epsfysize=75mm
 \centerline{
 \epsfbox{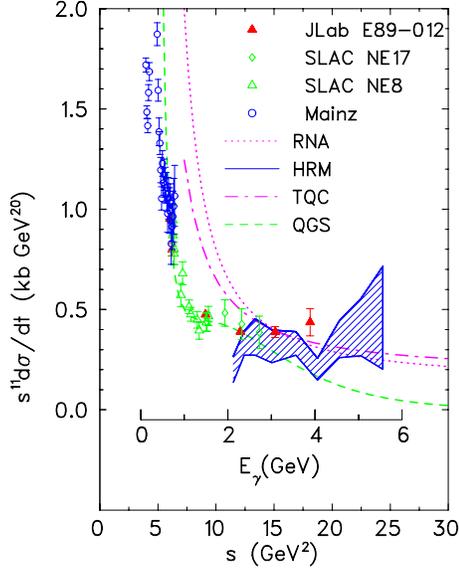}}
 \caption{
 The invariant cross section of the reaction
$\gamma d\to pn$ multiplied by $s^{11}$ 
 versus the value $s$ and  photon energy  $E_{\gamma}$.
 The figure is taken from Ref.\cite{brodsky2004}. 
 }
\label{f1brodsk}
\end{figure}

\section{The  CCR behaviour in exclusive 
 nuclear reactions }
 
 Few exclusive nuclear reactions were found to be compatible  with the CCR.
 So, the deuteron  photodisintegration reaction $\gamma d\to pn$
 follows the $s^{-11}$ scaling behaviour
 at photon energies
  $E_\gamma=1-4$ GeV and high transversal momenta
 $p_T>1.1 $ GeV/c corresponding to large scattering angles $\theta_{cm}\sim
90^\circ$
\cite{SLAC1} - \cite{rossi}
 (see Fig.\ref{f1brodsk}).  
 Meson-exchange models fail  to explain   the 
 $\gamma d\to pn$  data at $E_\gamma >1$ GeV (see, for example,
\cite{bochna}), and therefore several nonperturbative theoretical models 
 were suggested
\cite{nagorny}-\cite{grishina},\cite{MSU}. 
  Since the pQCD is expected to be valid at very high
 transferred momenta \cite{izgur}, the origin of the observed 
 scaling behaviour 
 in the reactions with the deuteron at moderate energies is unclear.
 On the other hand, in these reactions the
 3-momentum transfer $Q > 1$ GeV/c  is  large enough  to probe 
 very  short  distances between the nucleons   in nuclei,
 $r_{NN}\sim 1/Q<0.3 \,fm$.
 Most likely,
 nucleons could lose their separate identity
 in this overlapping region and form multi-quark configurations.
 In order to get more insight
 into  the underlying dynamics of the CCR behaviour,
 new data with the lightest nuclei are necessary.

\begin{figure}[ht]
\includegraphics[width=120mm,height=120mm]{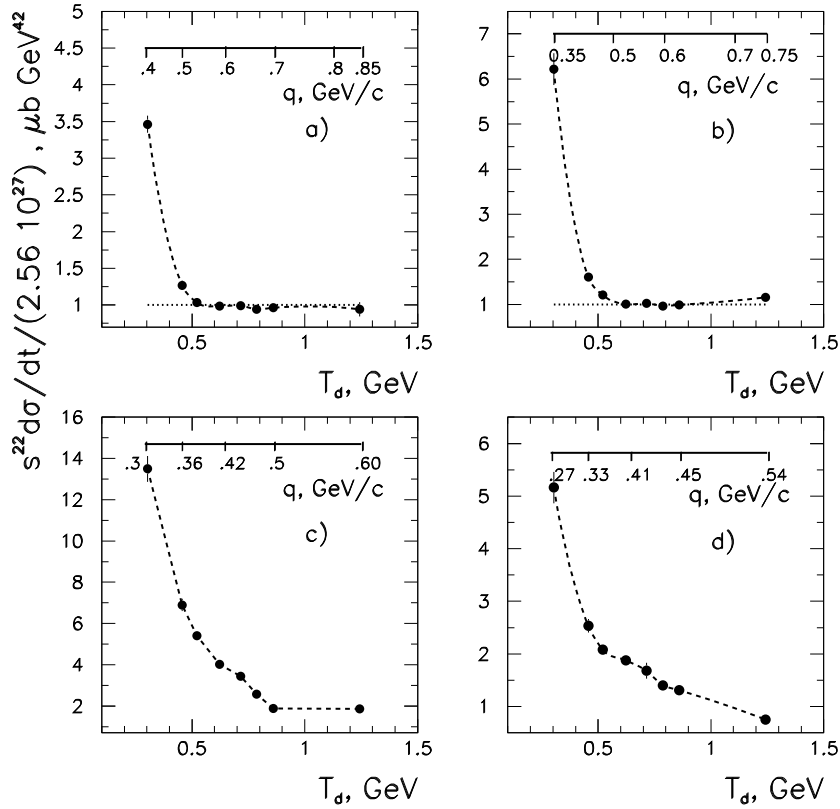}
 \caption{ 
The differential 
 cross section of the $dd \to~^3Hen$
 and $dd \to~^3Hp$  reactions
 multiplied by $s^{22}$  versus the deuteron beam energy at different
 scattering angles:
{\it a} -- $\theta_{c.m.}=60^o $; 
 {\it b} -- $50^\circ -52^\circ$;
{\it c} -- $33^\circ -35^\circ$; {\it d} -- $28^\circ$.
 On the upper scale is shown the minimal relative momentum 
 between nucleons in the deuteron for the ONE mechanism.
The data are taken from Ref. \protect\cite{bizard}.
 At lower scattering angles
  the plateu is not visible in
 this data, maybe, except for $\theta_{cm}= 33^\circ -35^\circ$.}
\label{s2260lin}
\end{figure}

\begin{figure}[hbt]
\epsfysize=120mm
 \centerline{
 \epsfbox{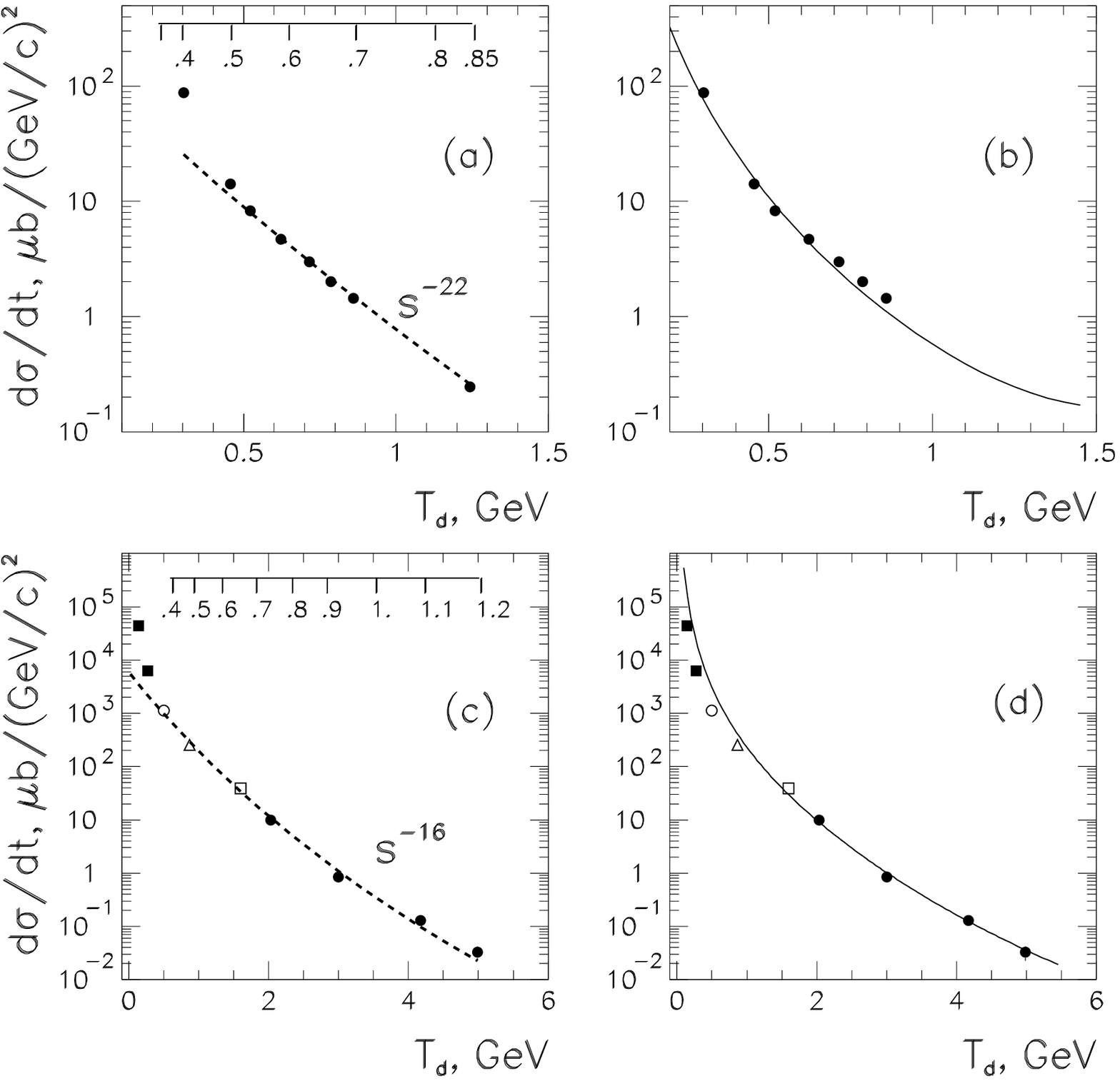}}
\caption
{
The differential 
 cross section of the $dd \to~^3He\,n$
 and $dd \to~^3H\,p$  reactions at $\theta_{c.m.}=60^o$  (a -- b)
 and $dp \to dp$  at $\theta_{c.m.}=127^o $ (c -- d)
 versus the deuteron beam kinetic energy.
 Experimental data in  (a -- b) ($\bullet $) are
 taken from \protect\cite{bizard}.
 In  (c -- d), the  experimental data (black squares),($\circ$), ($\triangle$),
(open square) and
($\bullet $) are taken from 
\cite{sekiguchi}-\cite{dubal}, 
respectively. 
  The dashed curves give the $s^{-22}$ (a) and $s^{-16}$ (c) behaviour.
  The full curves show the result 
  of calculations obtained in Ref. \cite{uz2005} using the Regge formalism. 
%
 On the upper scales in (a) and (c) is shown the minimum 
 relative momentum (GeV/c) 
 between nucleons in the deuteron for the ONE mechanism.
}
\label{fig2}
\end{figure}

 Vey recently was  shown \cite{uz2005} that the cross section
 of the reaction
 $dd\to ~^3Hen$ (and $dd\to ~^3H p$), measured at
 SATURNE in 80's \cite{bizard}, also perfectly follows the scaling behaviour 
 at transversal momenta $p_T\sim 0.6-0.9$ GeV/c (Fig.\ref{s2260lin}).
 In this reaction one has $n=6+6+9+3=24$. 
 At the beam energy  $T_d=0.5-1.25$ GeV the differential cross
 section $d\sigma/dt$  demonstrates   the $s^{-22}$ dependence
for the maximum measured
 scattering  angles $\theta_{cm}=50^\circ$ 
($\chi^2_{n.d.f.}=0.97$ for $T_d=0.62-1.24$ GeV) and $\theta_{cm}=60^\circ$ 
($\chi_{n.d.f.}^2= 1.18$ at $T_d=0.5-1.24$ MeV).
  {\it Up to now, the reaction $dd\to ~^3Hen$ ($~^3Hp$) is the only pure
 hadronic process which involves
  the deuteron and $^3He(^3H)$ nuclei and  found to follow the CCR.}
 As shown in \cite{uz2005}, the cross section of
 the reaction $dp\to dp$ also demonstrates the CCR behaviour
 $\sim s^{-16}$   at $T_d= 1-5$ GeV and 
 $\theta_{cm}= 120^\circ-130^\circ$ (Fig.\ref{fig2}), however the
 $\chi^2$-value is not good in this case, perhaps, due to different sets
 of the data included into analysis \cite{uz2005}.
 For other reactions with the lightest nuclei experimental
 data 
 in the  GeV  region and large scattering angles are either
 noncomplete, as for $d^3He\to ~^4He p$ and  $pd\to ^3H \pi^+$,
 or were obtained in different set of  experiments, what
 could lead to systematic uncertainties.

 In  Ref. \cite{Leksin} the reaction $dd\to ~^3Hen$ at high $p_T$
 is considered as a double cumulative process with nucleon cumulations
 involved  in both initial deuterons.
 From the point of view of constituent quark model the observed
 $s^{-22}$ behaviour implies
 that all constituent quarks in the initial and final state are 
 active in the $dd\to ~^3Hen$ ($~^3Hp$) reaction. 
 One should note that the nuclear matter density in the short-range
 configurations with high internal momenta between nucleons $q\sim 1 GeV/c$
 probed in this reaction, 
 is close to the critical one, $\varepsilon_c\sim 1GeV/fm^3$,
 that corresponds to the phase transition in cold baryon matter
 \cite{emeliyanov} 
 \footnote{ Of course, if the equilibrium thermodynamical state
  is formated on the  intermediate step of this reaction,
 then its  tepmerature 
 is  rather small at beam 
 energies $T_d\sim 1$ GeV: $kT_B \approx $ 50 MeV for baryon matter, 
  and $kT_q\approx  5$ MeV for constituent
 quark matter. This is much smaller than  the
 critical temperature $kT_c=150-170$ MeV expected  for  transition to
 quark-gluon plasma \cite{emeliyanov}.}.
  On the whole, interpretation of such phenomena  can be associated
 with the {\it quark-hadron duality}. In Ref.\cite{uz2005}
 the reggeon exchange  model \cite{grishina} was applied to
 the $dd\to ~^3Hen$ reaction  to clarify a possible
 relation to the $\gamma d\to pn$ data. One can see  from Fig.\ref{fig2}
 this model allows one to describe the data on the $dd\to ~^3Hp$
 and $dp\to dp$ reactions at high $p_T$.

\begin{figure}[hbt]
\epsfysize=90mm
 \centerline{\epsfbox{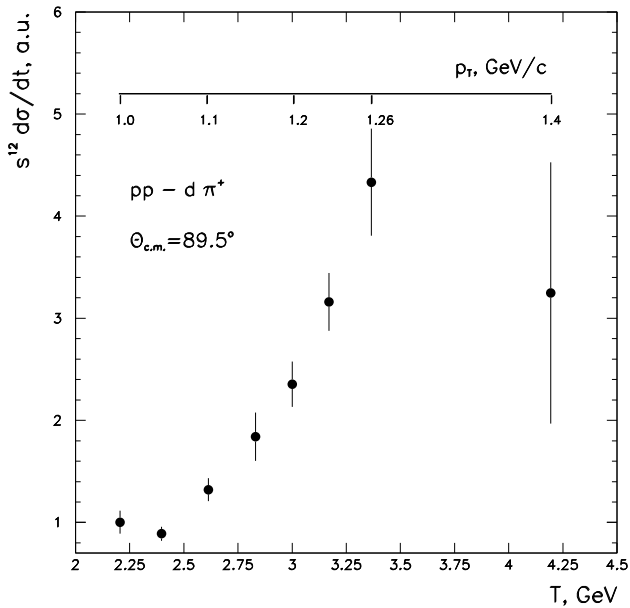}}
\caption
  {The cross section of the reaction  $pp\to d\pi^+$ $d\sigma/ dt$ 
 at  $\theta_{cm}=89.5^o$
 taken  from Ref.\cite{andersen} 
 multiplied by $s^{12}$ in arbitrary units as a function of beam energy.
 On the upper scale is shown the transversal momentum $p_T$.
 The scaling behaviour $s^{-12}$ is not observed.} 
\label{andersen74}
\end{figure}

 An important task addressed to experiment is a search for similar
 behaviour  in other exclusive reactions with the lightest nuclei
 at large $p_T\, (>0.6$ GeV/c).
 This task becomes more important in view of the fact that
 in the  reaction   $pp\to d\pi^+$, 
 measured   at the beam kinetic energy
 2.2 - 4.0 GeV  and large  angles of the pion production
 $\theta_{cm}=  40^\circ -90^\circ$  \cite{andersen}, 
 the CCR behavour was not observed (see Fig.\ref{andersen74}).
 This is in contrast with   the behaviour of the $\gamma d\to pn$  data 
 \cite{SLAC1}-\cite{rossi},
  although  the transversal momenta at these conditions in the inverse 
 reaction  $\pi^+d\to pp$
 are almost the same ($p_T > 1$ GeV/c) as in the
 $\gamma d\to pn$ reaction  in the observed  CCR region.
 Therefore, it would be interesting to check wether the reaction $pp\to
 d\rho^+$, which can be related to the $pn\to d\gamma$ by the vector 
 meson dominance,  follows the CCR.
 
\section{Conclusion}
 The scaling behaviour $s^{-11}$ of the cross section of the reaction
 $\gamma d\to pn$ starts at $p_T>1.1$ GeV/c and $E_{\gamma}> 1 $GeV 
\cite{rossi}, 
 whereas in the reaction $pp\to d\pi^+$
 the expected scaling regime $s^{-12}$ is not observed at the same $p_T$
\cite{andersen}. On the other hand, in the reaction
 $dd\to ~^3Hp$ the $s^{-22}$ behaviour is observed in the data of
 Ref.\cite{bizard} at lower transversal momenta  $p_T>0.6 $ GeV/c
 for $T_d>0.5-1.2$ GeV. Therefore, new data are required to get more
 insight on the origin of this scaling.

 Thus, boundaries
 of the scaling region for the $dd\to ~^3Hp$ reaction are not yet determined.
 As it seen from Fig.\ref{s2260lin}, within the ONE mechanism 
the CCR behaviour in the reaction
 $dd\to ~^3H\,p$ starts at internal momenta in the deuteron $q>0.5$ GeV/c
\footnote{In the $dp\to dp$ process
 the $s^{-16}$ behaviour also starts at $q>0.5$ GeV, as it seen from
 Fig.\ref{fig2}.}.
 It is important to verify this observation at other kinematical
 conditions, i.e. at higher beam energies and lower angles or at lower energies
 and higher angles.
 Furthermore,   new data  are necessary 
  to   check  whether the CCR behaviour
  is valid in the reaction $dd\to ~^3H\,p$  at higher energies $T_d>1.2$ GeV
 and larger scattering angles $\theta_{cm}=60^\circ -90^\circ$, corresponding
 to large relative momenta in the deuteron $q_{pn}>0.8$ GeV/c.
 On the whole, one has to know what is a true parameter for the
 scaling regime -- either internal nucleon momentum $q$, or the transversal
 momentum $p_T$.
  Experimental data  
 on other exclusive  reactions with the lightest nuclei
 where the baryon exchange mechanism are required: 
  $dd \to dd$, $pp\to \{pp\}_s\gamma$, $pd\to~^3H\pi^+$, $pd\to ~^3He\eta$,
 $dd\to ~^4He\eta$, $pp\to d\rho^+$, $d^3He\to~^4He\,p$.

{\bf Acknowledgments.} I am thankful to S.~Brodsky,
 A.B.~Kaidalov and  O.V.~Teryaev
  for useful discussion.

\end{document}